\documentclass[11pt,a4paper]{article}

\usepackage{epsfig,amsmath,amssymb,cite}

\begin{document}

\title{Gravitational lensing by a regular black hole}
\author{Ernesto F. Eiroa$^{1,2,}$\thanks{
e-mail: eiroa@iafe.uba.ar}, Carlos M. Sendra$^{1,2,}$\thanks{
e-mail: cmsendra@iafe.uba.ar} \\
{\small $^1$ Instituto de Astronom\'{\i}a y F\'{\i}sica del Espacio, C.C. 67, Suc. 28, 1428, Buenos Aires, Argentina}\\
{\small $^2$ Departamento de F\'{\i}sica, Facultad de Ciencias Exactas y 
Naturales,} \\ 
{\small Universidad de Buenos Aires, Ciudad Universitaria Pab. I, 1428, 
Buenos Aires, Argentina} }
\maketitle
\date{}

\begin{abstract}
In this paper, we study a regular Bardeen black hole as a gravitational lens. We find the strong deflection limit for the deflection angle, from which we obtain the positions and magnifications of the relativistic images. As an example, we apply the results to the particular case of the supermassive black hole at the center of our galaxy.
\end{abstract}



\section{Introduction}

The subject of gravitational lensing by black holes has received great attention in the last decade, basically due to the strong evidence about the presence of supermassive black holes at the center of galaxies, including ours \cite{ guillessen}. The study of black hole gravitational lenses can be simplified by using the strong deflection limit, which is an approximate analytical method for obtaining the positions, magnifications, and time delays of the images. It was introduced by Darwin \cite{darwin} for the Schwarzschild geometry, rediscovered by several authors \cite{otros}, extended to Reissner-Nordstr\"om geometry \cite{eiroto}, and to general spherically symmetric black holes \cite{bozza}. Kerr black holes were also analyzed adopting the strong deflection limit \cite{bozza1,bozza2,vazquez}. Numerical studies \cite{virbellis,virbhadra} were done too. Another related aspect that was considered, with the intention of measuring the properties of astrophysical black holes, are the shadows cast by rotating ones \cite{bhs,chandra}, which present an optical deformation caused by the spin, instead of being circles as in the non-rotating case. This subject have been recently re-examined by several authors \cite{shadow,bozza2} in the belief that the direct observation of black holes will be possible in the next years.  Black holes coming from alternative theories \cite{alternative}, braneworld cosmologies \cite{braneworld} and even naked singularities \cite{nakedsing} were considered as lenses. In the recent review article by Bozza \cite{bozzareview}, these topics are discussed with more detail, including the observational prospects and additional references. \\

The problem of singularities in general relativity, where the theory breaks down, is a long standing one, involving both black holes and cosmological solutions. Regular black holes are solutions of the equations of gravity for which an event horizon is present, but the spacetime is free from singularities. The first regular black hole was introduced by Bardeen \cite{bardeen}. This solution has both an event and a Cauchy horizons, but with a regular center, and it was shown by Borde \cite{borde} that the absence of the singularity is related to topology change. In Bardeen model the singularity is replaced by a de Sitter core. The Bardeen black hole can be interpreted as a magnetic solution of the Einstein equations coupled to nonlinear electrodynamics \cite{eloy}. Other spherically symmetric black holes with a regular center were studied by several authors, mainly in the context of non-linear electrodynamics \cite{regular} (for a recent review see Ref. \cite{ansoldi}).\\

In this article, we study gravitational lensing by a regular Bardeen black hole, in the strong deflection limit. The paper is organized as follows: In Sec. 2 we find the strong deflection limit approximation for the deflection angle; in Sec. 3 we obtain the positions and magnifications of the relativistic images and we calculate the observables corresponding to the supermassive Galactic black hole; finally, in Sec 4 we summarize the results obtained.

\section{Deflection angle}

Bardeen black holes are described by a spherically symmetric metric of the form
\begin{equation}
ds^{2}=-A(r)dt^{2}+B(r)dr^{2}+C(r)(d\theta^{2}+\sin^{2}\theta d\phi^{2}),
\label{m1}
\end{equation}
where the metric functions, working in geometrized units (speed of light in vacuum c=1 and gravitational constant G=1), are given by \cite{bardeen,borde,eloy}
\begin{equation}
A(r)=B^{-1}(r)=1-\frac{2Mr^{2}}{(r^{2}+g^{2})^{3/2}}, \hspace{0.5cm} C(r)=r^2,
\label{mc1}
\end{equation}
with $M$ being the mass and $g$ a constant, which can be interpreted as a charge in the context of nonlinear electrodynamics \cite{eloy}. For $g=0$ we recover the Schwarzschild geometry, which has a singularity at its center. When $g\neq 0$ this metric is regular everywhere, with a de Sitter behavior for small values of the radial coordinate $r$, and an asymptotically Schwarzschild behavior for large $r$. It is convenient to adimensionalize the metric (\ref{m1}) in terms of the Schwarzschild radius $2M$ by defining $x=r/2M$, $T=t/2M$, $q=g/2M$, so it becomes
\begin{equation}
ds^{2}=-A(x)dT^{2}+B(x)dx^{2}+C(x)(d\theta^{2}+\sin^{2}\theta d\phi^{2}),  
\label{m2}
\end{equation}
with
\begin{equation}
A(x)=B^{-1}(x)=1-\frac{x^{2}}{(x^{2}+q^{2})^{3/2}}, \hspace{0.5cm} C(x)=x^{2}.
\end{equation}
It was shown \cite{borde} that the existence of an event horizon for this metric is possible only in the case that $|q|\leq2/3\sqrt{3}$. For the strict inequality there are two horizons: the inner and the event ones; and  when $|q|=2/3\sqrt{3}$, the horizons fuse into one. Equating $A(x)=0$, we obtain
\begin{equation}
q^{6}+3q^{4}x^{2}+\left(3q^{2}-1\right)x^{4}+x^{6}=0, 
\label{eh}
\end{equation}
which is a third degree polynomial equation in $x^2$. Its largest positive solution corresponds to the event horizon radius $x_h$.\\
The photon sphere radius $x_{ps}$ is given by the largest positive solution of the equation \cite{weinberg}
\begin{equation}
\frac{A'(x)}{A(x)}=\frac{C'(x)}{C(x)}, 
\label{psg}
\end{equation}
where the prime represents the derivative with respect to $x$. For the Bardeen metric this equation can be written in the form of a fifth degree polynomial equation in $x^2$
\begin{equation}
-9 x^{8}+4 \left( q^2+x^2 \right) ^{5}=0, 
\label{ps}
\end{equation}
from which the value of $x_{ps}$ is obtained by using standard software.

The deflection angle for a photon coming from infinity, as a function of the closest approach distance $x_0$, has the form \cite{weinberg}
\begin{equation}
\alpha(x_0)=I(x_0)-\pi, 
\label{alfa1}
\end{equation}
where 
\begin{equation}
I(x_0)=\int^{\infty}_{x_0}\frac{2\sqrt{B(x)}dx}{\sqrt{C(x)}\sqrt{A(x_0)C(x)\left[ A(x)C(x_0)\right]^{-1}-1}}. 
\label{i0}
\end{equation}
The deflection angle grows if $x_0$ approaches to $x_{ps}$, where it diverges. Following Ref. \cite{bozza}, we define the new variable $z$ by
\begin{equation}
z=\frac{A(x)-A(x_0)}{1-A(x_0)}, 
\label{defz}
\end{equation}
and the functions
\begin{equation}
R(z,x_0)=\frac{2\sqrt{A(x)B(x)}}{A'(x) C(x)}\left[ 1-A(x_0) \right] \sqrt{C(x_0)}, 
\label{rz}
\end{equation}
\begin{equation}
f(z,x_0)=\frac{1}{\sqrt{A(x_0)-\left[\left(1-A(x_0)\right)z+A(x_0)\right]C(x_0) [C(x)]^{-1}}}, 
\label{fz}
\end{equation}
where $x=A^{-1}[(1-A(x_0))z+A(x_0)]$. By performing a Taylor expansion of the function inside the square root in Eq. (\ref{fz}) one obtains
\begin{equation}
f_0(z,x_0)=\frac{1}{\sqrt{\varphi (x_0) z+\gamma (x_0) z^{2}}}, 
\label{f0}
\end{equation}
where
\begin{equation}
\varphi (x_0)=\frac{1-A(x_0)}{A'(x_0) C(x_0)}\left[ A(x_0) C'(x_0) - A'(x_0) C(x_0)\right],
\label{varphi}
\end{equation}
and
\begin{eqnarray}
\gamma (x_0) &=& \frac{\left[ 1-A(x_0)\right] ^{2}}{2[A'(x_0)]^{3} [C(x_0)]^{2}}\left\{ 2 [A'(x_0)]^{2} C(x_0) C'(x_0) - A(x_0) A''(x_0) C(x_0) C'(x_0) \right. \nonumber \\
&& \left. + A(x_0) A'(x_0) \left[ C(x_0) C''(x_0) -2 [C'(x_0)]^{2}\right] \right\} .
\label{gamma}
\end{eqnarray}
With these definitions, the integral in Eq. (\ref{i0}) can be separated into two parts
\begin{equation}
I(x_0)=I_D(x_0)+I_R(x_0), 
\label{i0n}
\end{equation}
where 
\begin{equation}
I_D(x_0)=\int^{1}_{0}R(0,x_{ps})f_0(z,x_0)dz, 
\label{id}
\end{equation}
and
\begin{equation}
I_R(x_0)=\int^{1}_{0}[R(z,x_0)f(z,x_0)-R(0,x_{ps})f_0(z,x_0)]dz. 
\label{ir}
\end{equation}
If $x_0 \neq x_{ps}$ we have that $\varphi \neq 0$ and $f_0\sim 1/\sqrt{z}$, so the integral $I_D(x_0)$ converges. Instead, when $x_0= x_{ps}$, by using Eq. (\ref{psg}) we have that $\varphi =0$ and $f_0\sim 1/z$, so $I_D(x_0)$ has a logarithmic divergence. Therefore, $I_D(x_0)$ is the term containing the divergence at $x_0=x_{ps}$ and $I_R(x_0)$ is regular because it has the divergence subtracted. \\
\begin{figure}[t]
\centering
\includegraphics[width=16cm]{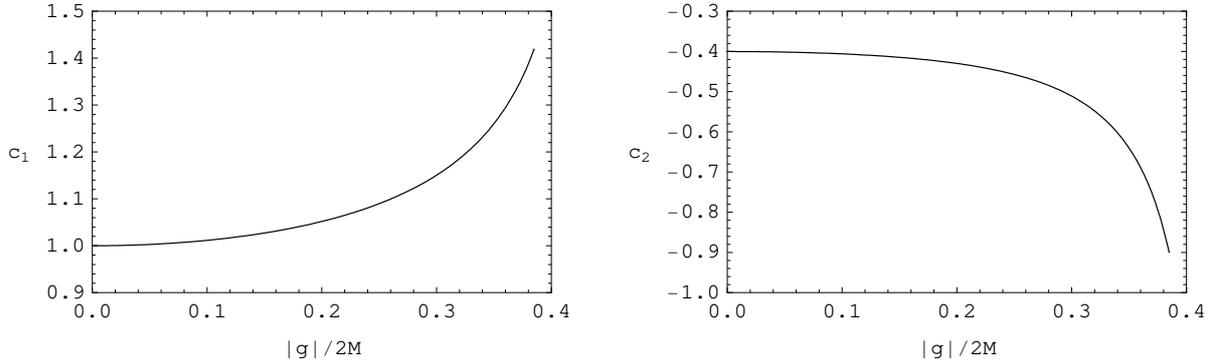}
\caption{The strong deflection limit coefficients $c_{1}$ (left) and $c_{2}$ (right) as functions of $|g|/2M$.}
\end{figure}
\begin{figure}[ht]
\vspace{0.5cm}
\centering
\includegraphics[width=8.5cm]{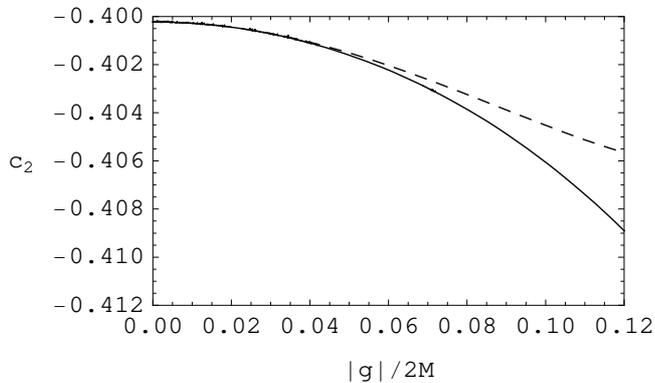}
\caption{Comparison of the strong deflection limit coefficient $c_{2}$ obtained numerically (full line) and to second order in $|g|/2M$ (dashed line). It can be seen from the plot that the Taylor approximation is good for small values of $|g|/2M$, as expected.}
\end{figure}
\begin{figure}[ht]
\centering
\includegraphics[width=16cm]{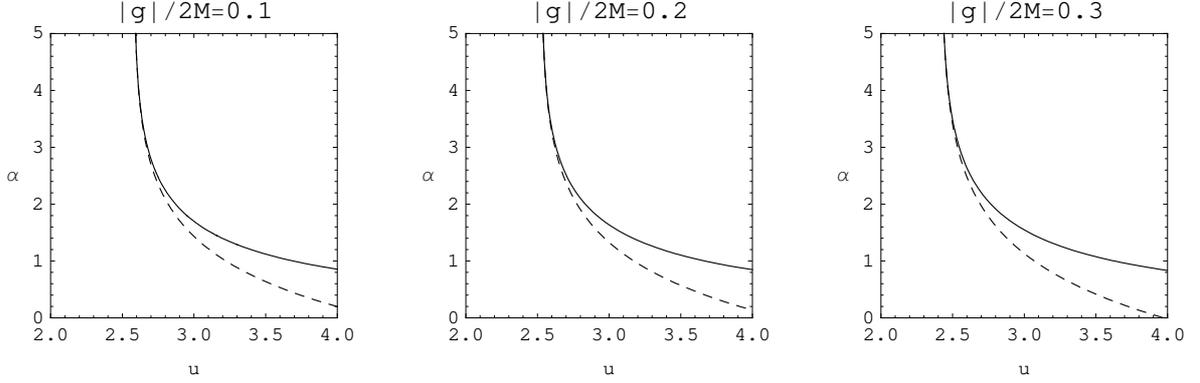}
\caption{Comparison of the exact deflection angle $\alpha$ calculated numerically (full line) and the corresponding strong deflection limit values (dashed line) as functions of the impact parameter $u$, for some values of $|g|/2M$.}
\end{figure}
The impact parameter $u$ is more easily connected to the lensing parameters than $x_0$. As a consequence of the conservation of angular momentum, $u$ is related to $x_0$ by 
\begin{equation}
u=\sqrt{\frac{C(x_0)}{A(x_0)}}. 
\label{u}
\end{equation}
Bozza \cite{bozza} showed that the logarithmic divergence of the deflection angle for photons passing close to the photon sphere has the form
\begin{equation}
\alpha(u)=-c_{1}\ln\left(\frac{u}{u_{ps}}-1\right)+c_{2}+O(u-u_{ps}), 
\label{ad}
\end{equation}
where $u_{ps}$ is the impact parameter evaluated at $x_0=x_{ps}$,
\begin{equation}
c_{1}=\frac{R(0,x_{ps})}{2\sqrt{\gamma (x_{ps})}}, 
\end{equation}
and
\begin{equation}
c_{2}=-\pi+c_R+c_{1}\ln\frac{2\gamma (x_{ps})}{A(x_{ps})}, 
\end{equation}
with
\begin{equation}
c_R=I_R(x_{ps}). 
\label{cr}
\end{equation}
This is the so-called strong deflection limit. For the Bardeen black hole we obtain that the strong deflection limit coefficients are given by
\begin{equation}
c_{1}=\frac{\left(q^{2}+x^{2}_{ps}\right)\sqrt{-2q^{2}+x^{2}_{ps}}}{\sqrt{8q^{6}-9q^{2}x^{4}_{ps}+x^{6}_{ps}\left[-1+3 \left( \sqrt{q^{2}+x^{2}_{ps}}\right) ^{-1}\right] }}, 
\end{equation}
and
\begin{equation}
c_{2}=-\pi+c_R+c_{1} \ln 2\delta, 
\end{equation}
where
\begin{equation}
\delta=\frac{ 8q^{6}-9q^{2}x^{4}_{ps}+x^{6}_{ps}\left[ -1+3 \left( q^{2}+x^{2}_{ps}\right) ^{-1/2}\right] }{\left(-2q^{2}+x^{2}_{ps}\right)^{3}\left[ 1-x^{2}_{ps}\left(q^{2}+x^{2}_{ps}\right)^{-3/2}\right] }. 
\label{delta}
\end{equation}
In our case, $c_R$ cannot be calculated analytically. Therefore, for small values of $|q|$, a second order power expansion is suitable as a good approximation:
\begin{equation}
c_R=c_{R,0}+c_{R,2}q^2+O(q^4), 
\label{cr2o}
\end{equation}
where $c_{R,0}$ is the value of $c_R$ for a Schwarzschild black hole and $c_{R,2}$ is the correction provided by the parameter $q$. We obtain
\begin{equation}
c_{R,0}=\ln\left[36\left(7-4\sqrt{3}\right)\right]\approx 0.949603, 
\label{cr0}
\end{equation}
\begin{equation}
c_{R,2}=\frac{4}{9}\left\{-18+4\sqrt{3}-5\ln\left[\frac{1}{6}\left(2+\sqrt{3}\right)\right]\right\}\approx -3.86568. 
\label{cr2}
\end{equation}
For large values of $|q|$, this Taylor expansion is not accurate, and $c_R$ should be obtained numerically. The numerical evaluation of the integral that gives $c_R$ is rather cumbersome, and $c_{2}$ can be obtained more easily in a similar way as it was done in Ref. \cite{eiroto}, from the limit
\begin{equation}
c_{2}=-c_{1}\ln \left[ \lim_{u\rightarrow u_{ps}}
\frac{u_{ps}e^{-\alpha (u)/c_{1}} }{u-u_{ps}} \right] .
\label{barblim}
\end{equation}
The strong deflection limit coefficients as functions of $q=g/2M$, with $c_{2}$ obtained numerically from Eq. (\ref{barblim}), are presented in Fig. 1. For small values of $|q|$ the Taylor expansion of $c_{2}$ in Eq. (\ref{cr2o}) gives good results, as shown in Fig. 2. For photons passing close to the photon sphere the strong deflection limit is an excellent approximation, as it can be seen in Fig. 3.

\section{Relativistic images}

We consider a black hole lens situated between a point source of light and the observer. The Bardeen space-time is asymptotically flat, and we take both the observer and the source far enough from the black hole to be in the flat region. In this Section, we find the positions and magnifications of the relativistic images and we obtain the observables for the supermassive black hole at the center of our galaxy.

\subsection{Positions and magnifications}

The optical axis can be defined as the line joining the observer ($o$) and the lens ($l$). The angular positions of the source ($s$) and the images, seen from the observer, are $\beta $ (positive) and $\theta $, respectively. The observer-source ($d_{os}$), observer-lens ($d_{ol}$) and the lens-source ($d_{ls}$) distances (measured along the optical axis) are taken much greater than the horizon radius. In this approximation, the lens equation has the form \cite{bozzale}
\begin{equation}
\tan \beta =\frac{d_{ol}\sin \theta - d_{ls} \sin (\alpha -\theta)}{d_{os} \cos (\alpha -\theta)} .
\label{pm1}
\end{equation}
The lensing effects are more important when the objects are highly aligned, so we will only study this case, in which the angles $\beta $ and $\theta $ are small, and $\alpha $ is close to an even multiple of $\pi $. When $\beta \neq 0$ two weak deflection primary and secondary images, which will be not analyzed here, and two infinite sets of point relativistic images are obtained \cite{darwin,virbellis}. The first set of relativistic images have a deflection angle that can be written as $\alpha =2n\pi +\Delta \alpha _{n}$, with $n\in \mathbb{N}$ and $0<\Delta \alpha _{n}\ll 1$. In this approximation, the lens equation results \cite{bozza,bozzale}
\begin{equation}
\beta =\theta -\frac{d_{ls}}{d_{os}}\Delta \alpha _{n}.
\label{pm2}
\end{equation}
For the other set of images, $\alpha =-2n\pi -\Delta \alpha _{n}$, therefore $\Delta \alpha _{n}$ should be replaced by $-\Delta \alpha _{n}$ in Eq. (\ref{pm2}). From the lens geometry it is clear that $u=d_{ol}\sin \theta \approx d_{ol}\theta $, so the deflection angle given by Eq. (\ref{ad}) takes the form
\begin{equation}
\alpha (\theta )=-c_{1}\ln \left( \frac{d_{ol}\theta }{u_{ps}}-1 \right) 
+c_{2}. 
\label{pm4}
\end{equation}
Inverting Eq. (\ref{pm4}) to obtain $\theta $ in terms of $\alpha $ and performing a first order Taylor expansion around $\alpha =2n\pi $, the angular position of the $n$-th image is given by
\begin{equation}
\theta _{n}=\theta ^{0}_{n}-\zeta _{n}\Delta \alpha _{n},
\label{pm6}
\end{equation}
with
\begin{equation}
\theta ^{0}_{n}=\frac{u_{ps}}{d_{ol}}\left[ 1+e^{(c_{2}-2n\pi )/c_{1}}
 \right] ,
\label{pm7}
\end{equation}
and
\begin{equation}
\zeta _{n}=\frac{u_{ps}}{c_{1}d_{ol}}e^{(c_{2}-2n\pi )/c_{1}}.
\label{pm8}
\end{equation}
From Eq. (\ref{pm2}), one has that $\Delta \alpha _{n}=(\theta _{n}-\beta ) d_{os}/d_{ls}$, and replacing it in Eq. (\ref{pm6}) 
\begin{equation}
\theta _{n}=\theta ^{0}_{n}-\frac{\zeta _{n}d_{os}}{d_{ls}}(\theta _{n}-\beta );
\label{pm10}
\end{equation}
then, using that $0<\zeta _{n} d_{os}/d_{ls}\ll 1$ and keeping only the first order term in $\zeta _{n} d_{os}/d_{ls}$, it is easy to see that the angular positions of the images are
\begin{equation}
\theta _{n}=\theta ^{0}_{n}+\frac {\zeta _{n}d_{os}}{d_{ls}}(\beta -\theta ^{0}_{n}).
\label{pm14}
\end{equation}
The second term in Eq. (\ref{pm14}) is a small correction on 
$\theta ^{0}_{n}$, so all images lie very close to $\theta ^{0}_{n}$. With a similar treatment, the other set of relativistic images have angular 
positions 
\begin{equation}
\theta _{n}=-\theta ^{0}_{n}+\frac {\zeta _{n}d_{os}}{d_{ls}}(\beta +\theta ^{0}_{n}).
\label{pm15}
\end{equation}
When $\beta =0$ (perfect alignment) instead of point images we obtain an infinite sequence of Einstein rings with angular radii
\begin{equation}
\theta ^{E}_{n}=\left( 1-\frac {\zeta _{n}d_{os}}{d_{ls}}\right) \theta ^{0}_{n}.
\label{pm16}
\end{equation}

It is well known \cite{schneider} that gravitational lensing conserves surface brightness and the quotient of the solid angles subtended by the image and the source gives the magnification of the $n$-th image:
\begin{equation}
\mu _{n}=\left| \frac{\sin \beta }{\sin \theta _{n}}
\frac{d\beta }{d\theta _{n}}\right|^{-1},
\label{pm17}
\end{equation}
and, using that the angles are small and Eq. (\ref{pm14}), we have
\begin{equation}
\mu _{n}=\frac{1}{\beta}\left[ \theta ^{0}_{n}+
\frac {\zeta _{n}d_{os}}{d_{ls}}(\beta - \theta ^{0}_{n})\right]
\frac {\zeta _{n}d_{os}}{d_{ls}},
\label{pm18}
\end{equation}
which can be approximated to first order in $\zeta _{n}d_{os}/d_{ls}$ by
\begin{equation}
\mu _{n}=\frac{1}{\beta}\frac{\theta ^{0}_{n}\zeta _{n}d_{os}}{d_{ls}}.
\label{pm19}
\end{equation}
The same expression is obtained for the other set of relativistic images. The magnifications decrease exponentially with $n$, so the first image is the brightest one. The magnifications are proportional to $(u_{ps}/d_{ol})^{2}$, which is a very small factor, then the relativistic images are very faint unless $\beta $ is close to zero (nearly perfect alignment). The magnification becomes infinite if $\beta =0$, and the point source approximation is no longer valid in this case. 

\subsection{Observables for the supermassive Galactic black hole}

The outermost image is the first one, with angular position $\theta _{1}$, and the others approach to the limiting value
\begin{equation}
\theta_\infty=\frac{u_{ps}}{d_{ol}},
\label{eqti}
\end{equation}
as $n$ increases. When the outermost image can be resolved from the rest of them, the lensing observables defined by Bozza \cite{bozza}
\begin{equation}
s=\theta _{1}-\theta _{\infty }
\label{defs}
\end{equation}
and
\begin{equation}
r=\frac{\mu _{1}}{\sum\limits_{n=2}^{\infty }\mu _{n}},
\label{defr}
\end{equation}
are useful. The observable $s$ is the angular separation between the first image and the limiting value of the succession of images, and $r$ represents the ratio between the flux of the first image and the sum of the fluxes of the other images. By using the strong deflection limit and for high alignment, they can be written in the form
\begin{equation}
s=\theta_\infty e^{(c_{2}-2\pi )/ c_{1}}, 
\label{eqs}
\end{equation}
and
\begin{equation}
r=e^{2\pi/c_{1}}. 
\label{eqr}
\end{equation}
By measuring $\theta_{\infty }$, $s$ and $r$ and inverting 
Eqs. (\ref{eqs}) and (\ref{eqr}), the strong deflection limit coefficients $c_{1}$ and $c_{2}$ can be obtained and their values can be compared with those predicted by the theoretical models to identify the nature of the black hole lens.\\

The Galactic center black hole, modeled using the Bardeen metric, is analyzed here to provide a numerical example. The black hole parameters, taken from Ref. \cite{guillessen}, are the mass $M=4.31 \times 10^{6}M_{\odot }$  and the distance from the Earth $d_{ol}=8.33$ Kpcs. The observables, as functions of the ratio between the charge and the Schwarzschild radius, are shown in Fig. 4. We see that $\theta_\infty$ and $r$ lessen with $|g|/2M$ while $s$ grows with $|g|/2M$.

\begin{figure}[t!]
\centering
\includegraphics[width=16cm]{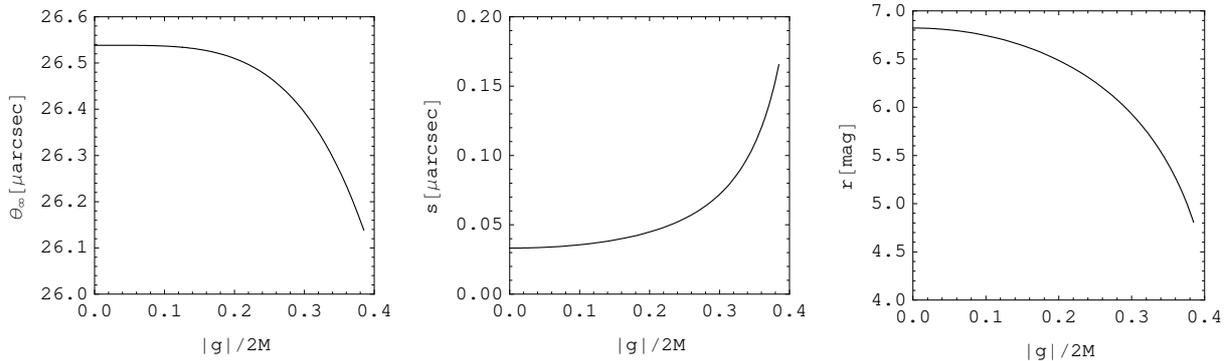}
\caption{Plots of the observables $\theta _\infty$ (left), $s$ (center), and $r$ in magnitudes [$r(mag)=2.5 log(r)$, right]  as functions of $|g|/2M$ for the supermassive Galactic black hole.}
\end{figure}

\section{Conclusions}

In this article we have obtained the strong deflection limit for the deflection angle corresponding to a regular Bardeen black hole, in terms of the self-gravitating monopole charge $g$ in non linear electrodynamics \cite{eloy}. This limit was calculated analytically by performing a second order Taylor expansion in the case of small values of the quotient $\left|g\right|/2M$ between the charge and twice the mass, while for any value it was found numerically. For close alignment, the resulting positions and magnifications of the relativistic images were obtained and the results were applied to calculate observables for the supermassive black hole situated at the center of our galaxy. We have found that when  $\left|g\right|/2M$ increases, the relativistic images are closer to the black hole. The angular separation $s$ also grows with $\left|g\right|/2M$, which means that the first relativistic image and the asymptotic limit of the angular positions of the images are more separated than in the case of the  Schwarzschild black hole. The first relativistic image also loses intensity with respect to the others if $\left|g\right|/2M$ increases. Then, for a Bardeen black hole, the first relativistic image is less intense with respect to the others than in the Schwarzschild geometry. Future instruments are expected to look into the region close to the Galactic supermassive black hole (see for example Ref. \cite{bozzareview} and references therein). Gravitational lensing would serve then as a tool to compare different black hole models with observations.

\section*{Acknowledgments}

This work has been supported by Universidad de Buenos Aires and CONICET.

\end{document}